\input harvmac
\noblackbox
\def\Title#1#2{\rightline{#1}\ifx\answ\bigans\nopagenumbers\pageno0\vskip1in
\else\pageno1\vskip.8in\fi \centerline{\titlefont #2}\vskip .5in}

\font\ticp=cmcsc10
%
%
\input epsf
\ifx\epsfbox\UnDeFiNeD\message{(NO epsf.tex, FIGURES WILL BE IGNORED)}
\def\figin#1{\vskip2in}
\else\message{(FIGURES WILL BE INCLUDED)}\def\figin#1{#1}
\fi
\def\Fig#1{Fig.~\the\figno\xdef#1{Fig.~\the\figno}\global\advance\figno
 by1}
%
%
%
%
\def\ifig#1#2#3#4{
\goodbreak\midinsert
\figin{\centerline{\epsfysize=#4truein\epsfbox{#3}}}
\narrower\narrower\noindent{\footnotefont
{\bf #1:}  #2\par}
\endinsert
}
%
%
\def\ajou#1&#2(#3){\ \sl#1\bf#2\rm(19#3)}
\def\jou#1&#2(#3){\unskip, \sl#1\bf#2\rm(19#3)}
\def\etc{{\it etc.}}
\def\zbar{{\bar z}}
\def\wbar{{\bar w}}

%
\lref\GrMe{D.J. Gross and P.F. Mende, ``The high-energy behavior of string scattering 
amplitudes,''\ajou  Phys. Lett. &197B (87) 129; ``String theory beyond the Planck 
scale,''\ajou  Nucl. Phys. &B303 (88) 407.}
\lref\Vene{D. Amati, M. Ciafaloni, and G. Veneziano, ``Superstring collisions at
 Planckian energies,''\ajou Phys. Lett. &197B (87) 81.}
\lref\PCJ{J. Polchinski, S. Chaudhuri, and C.V.
Johnson, ``Notes on D-branes,'' hep-th/9602052\semi J. Polchinski, ``TASI
lectures on D-branes,'' hep-th/9611050, 
to appear in the proceedings of the 1996
Theoretical Advanced Study Institute.}
\lref\Schw{Useful reviews are 
J.H. Schwarz, ``Lectures on superstring and M theory
dualities,'' hep-th/9607201, to appear in the proceedings of the 1996
Theoretical Advanced Study Institute\semi
M.R. Douglas, ``Superstring dualities, Dirichlet branes, and the small
scale structure of space,'' hep-th/9610041, to appear in the proceedings of
the Les Houches School on Theoretical Physics, Session 64, Quantum
Symmetries, Les Houches, France, 1995. }
\lref\Vafa{C. Vafa, talk at 1996 CERN Workshop on S-duality.}
\lref\BFSS{T. Banks, W. Fischler, S.H. Shenker, and L. Susskind, ``M theory
as a matrix model: a conjecture,''  hep-th/9610043.}
\lref\Shen{S.H. Shenker, ``Another length scale in string theory?''
hep-th/9509132.}
\lref\KlTh{I.R. Klebanov and L. Thorlacius, ``The size of p-branes,''
hep-th/9510200\jou Phys. Rev. Lett.& B371 (96) 51.}
\lref\Fish{W. Fischler, S. Paban, and M. Rozali, ``Collective coordinates
for D-branes,'' hep-th/9604014\jou Phys. Lett &B381 (96) 62 \semi D.
Berenstein, R. Corrado, W. Fischler, S. Paban, and M. Rozali, ``Virtual
D-branes,'' hep-th/9605168\jou Phys. Lett. &B384 (96) 93.}
\lref\Bach{C. Bachas, ``D-brane dynamics,'' hep-th/9511043\jou Phys. Lett.
&B374 (96) 37.}
\lref\Doug{M.R. Douglas, ``Gauge fields and D-branes,'' hep-th/9604198.}
\lref\PoSh{D. Kabat and P. Pouliot, ``A comment on zero-brane quantum
mechanics,''  hep-th/9603127\jou Phys. Rev. Lett. &77 (96) 1004\semi
M.R. Douglas, D. Kabat, P. Pouliot, and S.H. Shenker, ``D-branes and short
distances in string theory,'' hep-th/9608024.}
\lref\DDF{E. Del Giudice, P. DiVecchia, and S. Fubini, ``Factorization and
operator formalism in the generalized Veneziano model,'' \ajou Nuov. Cim. &
5A (71) 90.}
\lref\GSW{See e.g. M.B. Green, J.H. Schwarz and 
E. Witten, {\sl Superstring Theory}, Chapter 2.}
\lref\Bakl{V. Balasubramanian and I.R. Klebanov, ``Some aspects of massive
world brane dynamics,'' hep-th/9605174\jou Mod. Phys. Lett. &A11 (96) 2271.}
\lref\Mitu{D. Mitchell, N. Turok, R. Wilkinson, and P. Jetzer, ``The decay
of highly excited open strings,'' \ajou Nucl. Phys. &B315 (89) 1 \semi 
D. Mitchell, B. Sundborg, and N. Turok, ``Decays of massive open strings,''
\ajou Nucl. Phys. &B335 (90) 621. }
\lref\BaSu{T. Banks and L. Susskind, ``Brane -- anti-brane forces,''
hep-th/9511194.}
\lref\SGTasi{S.B. Giddings, ``Fundamental strings,'' in the proceedings of
the 1988 Theoretical Study Institute, A. Jevicki and C.-I Tan, eds. (World
Scientific, 1989).}

\Title{\vbox{\baselineskip12pt\hbox{UCSBTH-96-29}\hbox{hep-th/9612022}
}}
{\vbox{\centerline {Scattering Ripples from Branes}
}}
\centerline{{\ticp Steven B. Giddings}\footnote{$^\dagger$}
{Email address:
giddings@denali.physics.ucsb.edu}
}
\vskip.1in
\centerline{\sl Department of Physics}
\centerline{\sl University of California}
\centerline{\sl Santa Barbara, CA 93106-9530}

\bigskip
\centerline{\bf Abstract}

A novel probe of D-brane dynamics is via scattering of a high energy ripple
traveling along an attached string.  The inelastic processes in which the
D-brane is excited through emission of an additional attached string is
considered.  Corresponding amplitudes can be found by factorizing 
a one-loop amplitude derived in this paper.  This one-loop amplitude is
shown to have the correct structure, but extraction of explicit expressions
for the scattering amplitudes is difficult.  It is conjectured that the
exponential growth of available string states with energy leads to an
inclusive scattering rate that becomes large at the string scale, due to
excitation of the ``string halo,'' and meaning that such probes do not
easily see structure at shorter scales.

\Date{}

\newsec{Introduction}

Following the string revolution of ca. 1984, theoretical physics was in the
unique situation of having a candidate for the ``theory of everything,"
namely the heterotic string.  The ongoing string revolution of the 1990's
has changed that picture considerably, particularly as a result of the
discovery of the importance of D-brane states\foot{See \refs{\PCJ} for
recent reviews.} and dualities between
different vacua \refs{\Schw}.
One question that no longer apparently has a clear answer is that
of describing the fundamental degrees of freedom of the theory we are
studying.

In particular, the relative roles of D-brane states and string states are
uncertain.  Although D-branes seem to be solitons of string theory, it is not
obvious that they are solitons in precisely the same
sense as other more familiar
solitons in field theory.  A useful analogy is the role of
monopoles in QED versus that in a grand unified theory.  There is a
consistent set of rules, first introduced by Dirac, that provides a
treatment of many
of the physical phenomena in the presence of a monopole within
QED.  However, there are certain questions that don't have answers in QED,
such as the scattering behavior of Callan-Rubakov modes or the amplitude
for high-energy $e^+$--$e^-$ scattering to produce a monopole anti-monopole
pair.  In contrast, in grand unified theories, monopoles are true solitons
in the sense that they can be constructed smoothly from quanta of the
underlying gauge and Higgs fields.\foot{More precisely, widely 
separated monopole-antimonopole pairs
can be so created.}  This fact allows explicit treatment of both the
Callan-Rubakov modes and of monopole production, and all questions are in
principle answerable in terms of the Yang-Mills/Higgs dynamics of the fundamental
fields.  

It is far from clear that D-branes are solitons of string theory in the
same sense.  For one thing, D-branes carry fundamental charge, namely
Ramond-Ramond charge, not carried by the string states.  It is not 
immediately apparent
that a D-brane can be smoothly assembled from underlying string degrees of
freedom.  In fact, at least two alternatives have been proposed in the
literature.  The first\refs{\Vafa}
is that there is a sort of democracy, namely strings
are usefully treated as the fundamental degrees of 
freedom in some regions of the moduli space
of theories, and D-branes are most usefully taken as fundamental in other
regions.  The two descriptions would then be patched together at
intermediate moduli.  A second\refs{\BFSS} is that the D0-branes are
fundamental constituents of the theory from which other states can be
assembled.  

To give this question a sharp edge it is useful to turn to scattering
problems.  A crucial sharp question that doesn't appear to have a good
answer in the present formalism is that of computing the
amplitude for a high-energy annihilation of a pair of strings to
create a pair
consisting of a D0-brane and its antiparticle.  This type of problem may
well push the first alternative above past its limits, and drives
towards the heart of the question of what the fundamental degrees of
freedom are.\foot{Banks and Susskind \refs{\BaSu} raised the closely
related puzzle of what happens when a D0 pair annihilate, and give
arguments for the current intractability of this problem.}  

It would also be
useful to investigate other types of scattering phenomena to search for
other problems analogous to the pair production problem, or to
the Callan-Rubakov problem in QED, where the scattering
description for certain modes manifestly breaks down.  Furthermore, given
evidence for
structure at distances below the string scale \refs{\Shen},
a related problem is that of looking for scattering phenomena that
further exhibit
and illuminate such short-distance structure.

Various types of scattering phenomena involving branes have been
considered.  Refs.~\refs{\KlTh-\Fish} treated scattering of strings from
branes.  Except in the case of D-instantons, this type of scattering is
apparently dominated by dynamics at the string scale.  Brane-brane
scattering has been considered in \refs{\Bach\Doug-\PoSh}, and although it
has revealed evidence of distances at shorter scales, is somewhat difficult
to treat in generality. 

The purpose of this paper is to begin an investigation of another type of
scattering involving branes.\foot{This type of scattering was first
described in \refs{\PCJ}.}  Consider a long string stretched between two
widely separated branes.  One can attempt to probe the structure of one of
the branes by sending a ripple down the string to collide with the brane.
Furthermore, in \PCJ\ it was argued that such scattering might be capable
of revealing structure at sub-stringy scales.  To see this, consider
working at weak string coupling $g$.  To leading order, the ripple is
simply reflected due to the Dirichlet boundary conditions at the end of the
string.  This corresponds to a constant (frequency independent) phase
shift, and is indicative of point-like structure.  Before jumping to
conclusions, however, one should consider higher order processes such as
one where a ripple collides with a brane and produces an excitation of the
brane.  These processes first enter at order $g$ in perturbation theory.
Suppose for example that the inclusive amplitude for such processes grew as
$$
{\cal A}\sim g \omega^p\ ,
$$
for some $p$, at high frequency.  This amplitude only becomes substantial,
indicating possible breakdown of pointlike structure, at frequencies of
order $g^{-1/p}$.  This would indicate that such scattering is governed by
dynamics at shorter scales than the string scale.

This paper will begin an investigation of higher order processes with the
order $g$ amplitude.  This will be treated by applying the optical theorem
to, or equivalently factorizing, the appropriate one-loop amplitude.
Section two will
discuss the kinematical setup and related issues.  Section three derives an
expression for this amplitude, and the following section attempts an
analysis of this amplitude in the open-string factorization limit.
Unfortunately the resulting expressions have not been sufficiently
tractable to extract a definitive answer about the high-energy scattering
behavior.  However, as argued in section five, one plausibly discovers
high-energy amplitudes that grow exponentially in the energy.  This
behavior could result from the exponential growth of the number of
available final states.  If this is the correct behavior, it means that the
scattering becomes large at least before a scale only logarithmically
down from the string scale, and suggests that structure beyond the string
scale is difficult to probe by scattering ripples from branes.

The appendix collects some formulas useful for computing one-loop
amplitudes.
\newsec{Groundwork}

Our goal is to study scattering of a high-energy pulse, traveling along
a string, off of an attached D-brane. To simplify the calculation, I'll
begin by considering such a pulse traveling on a string stretched
between two parallel branes with separation ${L}$, and then take the
limit ${L}{\rightarrow}{\infty}$.

Creating a high frequency state on the string means creating a state
with a level number scaling like ${L}$ as ${L}{\rightarrow}{\infty}$. To
be precise, let $|{0},{L}{\rangle}$ be the vacuum in the sector of
strings connecting the two branes. Acting with ${\alpha}^\mu_{-n}$
produces an oscillation of frequency
\eqn\katie{
{\omega} = {{{\pi}{n}}\over{L}}.}
If we consider the mass-shell condition (with $\alpha'=1/2$)
\eqn\masssh{
{M^2} = {{L^2}\over{\pi^2}} + {2}\,({n} - {1})}

\noindent in the limit ${L}{\rightarrow}{\infty}$, we indeed see that the rest
frame energy is shifted by ${\omega}$:
\eqn\sally{
{E_{cm}} = {M} = {{L}\over{\pi}} + {\omega} + {\cal O}\,({{1}\over{L}}).}

However, acting with ${\alpha}^\mu_{-n}$ does not produce a physical
state satisfying the Virasoro conditions 
\eqn\tina{
({L_0}-{1})\,|{\rm phys}{\rangle} = {L_n}\,|{\rm phys}{\rangle} = {0}\quad
(n>0)\ .}
This raises the possibility of spurious behavior from the unphysical
part of the operator. This potential pitfall can be avoided by working
only with physical states, for example by working in the light-cone
gauge. With D-branes, however, there is a new subtlety. Light cone gauge
corresponds to picking out a null combination ${X^+} = {X^0} +
{X^1}$ and identifying it with world-sheet time, 
\eqn\joy{
{X^+}\,\,{\propto}\,\,{\tau}.}
In particular, note that this implies ${X^+}$ satisfies Neumann boundary
conditions
\eqn\denise{
{n^a}{\nabla_a}{X^+} = {0}}

\noindent on the world-sheet boundary. Therefore ${X^0}$ and ${X^1}$
can't be transverse to the brane, and our treatment applies only to p-branes
with ${p}{\geq}{1}$.\foot{Note that this adds a subtlety to the
usual approach of cataloging physical states, based on the light-cone
frame, for the case of strings attached to D-particles or D-instantons.}
Coordinates will therefore be taken to be ${X^i}$, transverse to the
brane, and ${X^I} = \lbrace{X^{\pm}}, {X^a}\rbrace$, tangent to the brane.

Rather than working directly in the light-cone gauge, I'll instead work
in the covariant framework using the DDF operators\refs{\DDF} to
translate the physical states from light cone gauge. To construct these,
let ${p_0}$ be a momentum satisfying the unexcited mass-shell condition 
\eqn\jane{
{p^2_0} = {-}\,\biggl({{L}\over{\pi}}\biggr)^2\,\,+\,\,{2}}
\noindent and choose a null vector ${k}$ (defining the directions $X^{\pm}$)
from the subspace tangent
to the brane, such that 
\eqn\hope{
k{\cdot}{p_0}\,\,=\,\,{1}.}
\noindent The DDF operators are then\foot{See \refs{\GSW} for
elaboration on the properties of DDF operators.} 
\eqn\ddfop{
\eqalign{
{A}^a_{-n}&= {{1}\over{2\pi}}\,\,{\int}^{2\pi}_0\,\,{d\tau}\,\,{\dot
X}^a\,\,{e^{-ink\cdot X}}\cr
{A}^i_{-n}&={{1}\over{2\pi}}\,\,{\int}^{2\pi}_0\,\,{d\tau}\,\,{X}^{i\prime}
{e^{-ink\cdot X}}\cr
}}
\noindent for oscillators parallel or transverse to the brane,
respectively. Here the integrals are carried out along the world-sheet
boundary, which we will assume is attached to one of the branes.

To summarize, the picture is as follows. We have a string attached to
two parallel branes, one through the origin and one at a distance ${L}$.
We act with a DDF operator ${A}^a_{-n}$ or ${A}^i_{-n}$ at the boundary
attached to the brane at ${L}$. This creates a plane wave-like oscillation of
the string of energy  ${\omega} = {\pi n}/{L}$ satisfying the Virasoro
conditions and the mass-shell condition \masssh.
We then wish to consider the limit ${L}{\rightarrow}{\infty}$,
holding
${\omega}$ fixed and large, and determine what final states are produced
in scattering the oscillation, or ripple, off the brane.

This is a difficult general problem so I'll specialize further. In
particular, this paper will focus on the one-string emission amplitude,
as this can be found by factorizing the one loop amplitude. For further
simplicity I'll also consider only polarizations parallel to the brane,
although many of the formulas translate directly to polarizations
perpendicular to both the brane and the direction separating the
branes.

\newsec{The One-Loop Amplitude.}

\ifig{\Fig\LoopD}{In this one-loop diagram, a string stretching between the
branes at $0$ and $L$ is created at point 1.  This is excited by a DDF
operator at 2, acting on the boundary of the string at $L$.  These states
are likewise annihilated at 3 and 4.  The upper boundary of the annulus is
fixed on the brane at $0$.  The annulus is taken to have circumference 1
and height $\lambda$.  Factorizing in the $\lambda\rightarrow 0$ 
limit, we find the amplitude for the
oscillation of the long string to create an excitation of the brane at $0$
involving a string with both its ends attached to the
brane.}{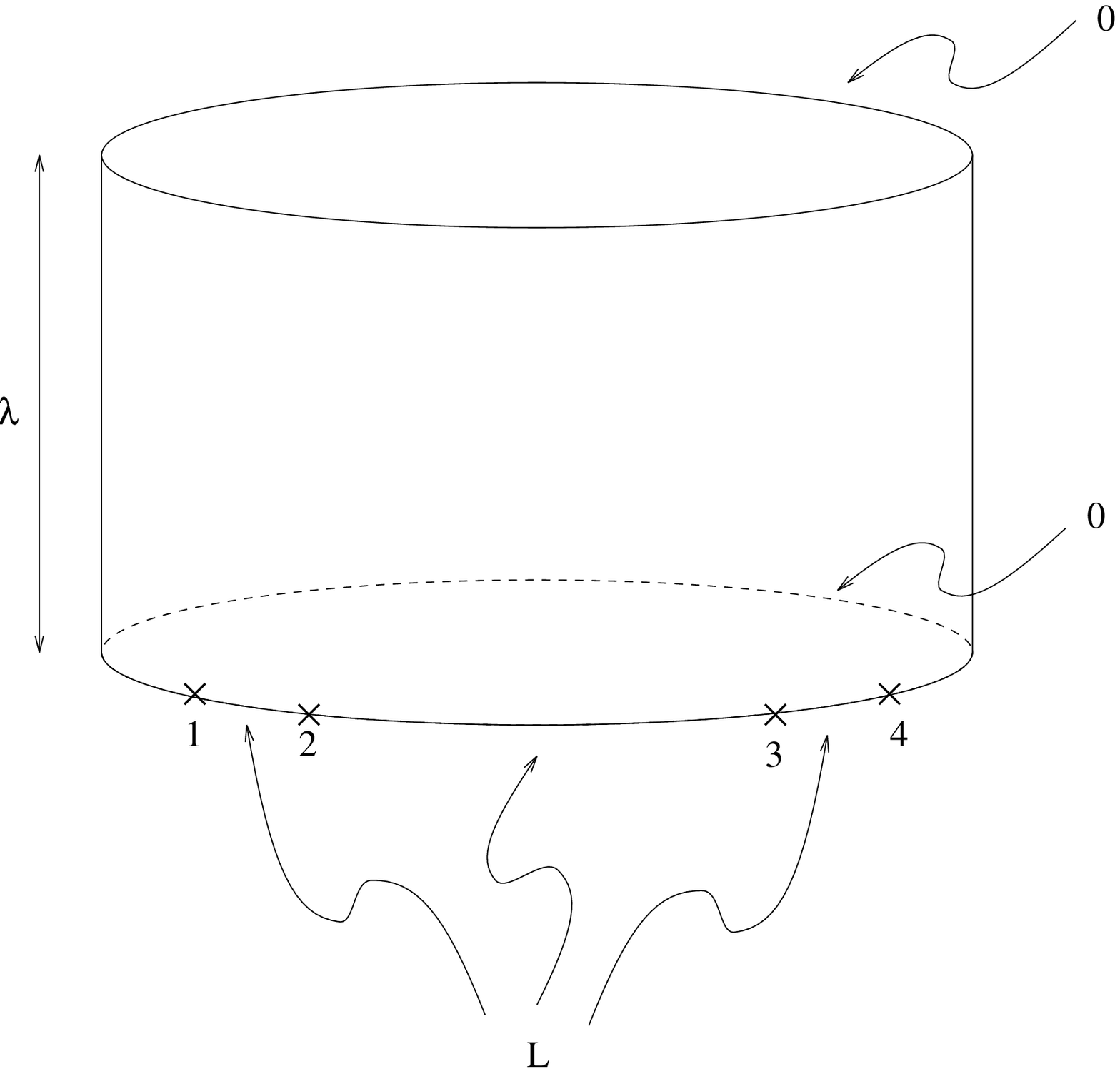}{3.25} 

As stated above, inclusive scattering with one final string can be
studied by factorizing the one-loop amplitude. To be precise, we
consider the four-point function on the annulus, as shown in fig. 1.
Two of the external states, created by operators ${\cal O}_1$ and ${\cal
O}_4$, correspond to the ground state of the string stretched between
the branes,
\eqn\lindy{
\eqalign{
{\cal O}_1&= {e^{ip_0\cdot X}}\,\,{\cal O}_D\,({\tau_1})\cr
{\cal O}_4&= {e^{-ip_0\cdot X}}\,\,{\cal O}_D\,({\tau_4}).\cr
}
}
\noindent Here ${\cal O}_D$ are operators that create string states with
the appropriate Dirichlet boundary conditions. The other two operators
are DDF operators (de)exciting the strings,
\eqn\meg{
\eqalign{
{\cal O}_2&= {A}^a_{-n}\cr
{\cal O}_3&= {A}^a_n\ .\cr
}
}
\noindent These operators may be taken arbitrarily close to ${\cal O}_1$
and ${\cal O}_4$, respectively. The processes 

\centerline{{\rm ripple + brane}${\rightarrow}$ {\rm ripple + excited brane}}

\noindent can be investigated by factorizing this amplitude in the open
string channel, ${\lambda}{\rightarrow}{0}.$  At intermediate times, the
excitation of the brane arises from an open string with both ends attached
to the brane.

The one-loop amplitude is given by the standard Polyakov formula,
\eqn\tootie{
{\cal A} = {\int}\,\,{d}{\mu}({m_i})\,\,{\int}\,\,{\cal D}{X^I}
{\cal D}{X^i}\,\,{e^{-S}}\,\,{\cal O}_1{\cal O}_2{\cal O}_3{\cal O}_4
}
\noindent where ${d}{\mu} ({m_i})$ is the measure for moduli and 
\eqn\doris{
{S} = {{1}\over{2\pi}}\,\,{\int}\,\,{d^2\sigma} {\sqrt{g}}\,\,({\nabla}
{X})^2\ .
}
The integral over
${X^\pm}$ is straightforward to evaluate by standard methods, and gives
\eqn\xpmint{
\biggl({{{det'_N}{\Delta}}\over{\int\,d^2\sigma{\sqrt{g}}}}\Biggr) ^{-1}\,\,
{\rm exp}\biggl\lbrace{{1\over4}{\sum_{i,j}} P_i \cdot P_j G_N
(\tau_i,\tau_j)}
\biggr\rbrace
}
 where the determinant is taken with Neumann boundary
conditions, prime denotes omission of the zero mode, 
${P_i}$ represent the four external state momenta, and
\eqn\helen{
{G_N}({\tau_i},{\tau_j}|{\lambda}) = {2}\,{\ell n}\,\,
\Biggl\lbrack{{{\theta_1}({\tau_i}-{\tau_j}|{2i}{\lambda})}\over
{{\theta_1}({\tau_i}-{\tau^{\prime}}|{2i}{\lambda})}}\Biggr\rbrack
}
\noindent is the Neumann Green function (see appendix). Eq.~\xpmint\
becomes
\eqn\xpminti{
{\int d^2\sigma \sqrt g\over det'_N\Delta}\,\,\Biggl
\lbrack{{{\theta_1}({\tau_1}-{\tau_3}
|{2i}{\lambda}){\theta_1}({\tau_2}-{\tau_4}|{2i}{\lambda})}\over
{{\theta_1}({\tau_1}-{\tau_2}|{2i}{\lambda}){\theta_1}({\tau_3}-{\tau_4}|
{2i}{\lambda})}}\Biggr\rbrack^n\,\,{\theta_1}({\tau_1-\tau_4}|{2i}{\lambda})
^{-p_0^2}{\theta_1}({\epsilon}|{2i}{\lambda})^{p_0^2}
}
\noindent where ${\epsilon}$ is a worldsheet UV regulator. Likewise the
integral over ${X^a}$ gives
\eqn\sara{\left( {det}'_N\,{\Delta}\over \int d^2\sigma \sqrt g
\right)^{-(p-1)/2}\,\,[{\ell n}{\theta_1}]^{\prime\prime}
({\tau_2}-{\tau_3}).
}

\indent The integral over ${X^i}$ involves the operators ${\cal O}_D$. We need
not treat these explicitly, but rather note that they create states
satisfying the Dirichlet boundary conditions
\eqn\faye{
\eqalign{
{X}&=0~~~~~{\tau}>{\tau_4}, {\tau}<{\tau_1}\cr
{X}&={L}~~~~~{\tau_4}>{\tau}>{\tau_1}\cr
}
}
\noindent in the coordinate separating the branes.

A general formula for the functional integral with Dirichlet boundary
conditions

\eqn\alice{
{X}{|}_{\partial} = x({\tau})
}
\noindent is

\eqn\robin{\eqalign{
\int_{X |_{\partial}= x(\tau) }& {\cal D}X\,e^{-S}=\cr &
{\int_{X|_\partial=0}}{\cal D}X\,e^{-S}
{\rm exp}\Biggl\lbrace-{{1}\over{4\pi^2}}
\,{\oint}\,{d\tau}\,{\oint}
{d}{\tau^{\prime}}\,\,{x}({\tau}){x}({\tau^{\prime}}) {\partial_n}
{\partial_{n^{\prime}}}{G_D}({\tau},{\tau^{\prime}})\Biggr\rbrace}}
\noindent where the normal derivatives ${\partial_n}$ act on the
Dirichlet Green function (see appendix)
\eqn\teresa{
{G_D}(z,w|\lambda) = {1\over2}{\ell n}\Biggl\lbrack{{{\theta_1}(z-w|2i\lambda)
{\theta_1}
({\bar z}-{\bar w}|{2i}\lambda)}\over
{{\theta_1}({z}-{\bar w}|{2i}\lambda){\theta_1}({\bar z}
-{w}|{2i}\lambda)}}\Biggr\rbrack\,\,+\,\,{{2\pi}\over{\lambda}}\,{\rm Im}
{\it w} {\rm Im} {\it z}.
}
\noindent From this we derive the result for the integral over ${X^a}$,
\eqn\jackie{
{det_D}{\Delta}^{-(d-p-1)/2}\,\,{\theta_1}({\tau_4}-{\tau_1}
|2i\lambda)
^{-L^2/\pi^2}\,\,{e^{-L^2 (\tau_1-\tau_4)^2/2\pi\lambda}}\,\,
{\theta_1}({\epsilon}|2i\lambda)\,^{L^2/\pi^2},
}
\noindent where the determinant is taken with Dirichlet boundary
conditions and $d=26$ for the bosonic string.
 
The determinants are readily evaluated (see
appendix), and give
\eqn\tammy{
\eqalign{
{{{det'_N}{\Delta}}\over{{\int}{d^2\sigma}{\sqrt{g}}}}&=
2{q^{1/12}}{f}({q^2})\cr
{det_D}{\Delta}&= 2{\lambda}{q^{1/12}}{f}({q^2})
}
} 
\noindent with ${f} ({q^2}) = {\prod_{n=1}^\infty} 
({1}-q^{2n})$ and  
${q} = {e^{-2\pi\lambda}}$,  
\noindent and these combine with the measure into the result
\eqn\karen{\eqalign{
{d\mu(m_i)}& \Biggl( {{det'_N\Delta}\over{\int d^2\sigma{\sqrt{g}}}}\Biggr)
^{-(p+1)/2}({det'_D}{\Delta})^{-(d-p-1)/2}\cr &=
{d\lambda}{d}{\tau_4}\,{\lambda}^{-(26-p-1)/2}2^{-12}{q^{-2}}{f^{-24}}
({q^2}).}
}
Finally, note that the integrals in the DDF operators \ddfop\ involve
the coordinate ${\tau}$ intrinsic to the infinite strip string world
sheet corresponding to string propagation. When this is continued to
euclidean space and mapped to a portion of the annulus, using ${z} =
{e^{i\tau + i\sigma}}$, with the incoming state mapped to a point, then
the integral is transformed to one over a coordinate ${z}$ that 
circles the incoming state:
\eqn\june{
{\int^{2\pi}_0}\,{d}{\tau}{\longrightarrow}{\oint_{z=z_0}}\,{dz}.
}
\indent Collecting these formulas, setting ${\tau_1}={0}$, renaming 
${\tau_4}={\tau}$, and
dropping an overall infinite factor, we arrive at the expression for the
desired one loop amplitude:
\eqn\famp{
{\cal A} = {\int^\infty_0}\,{d\lambda}\,{\int^{1}_0}\,{d}{\tau}\, 
{\lambda}^{-(26-p-1)/2}2^{-12}{q^{-2}}{f^{-24}}({q^2})\, 
\Biggl\lbrack{{{\theta_1}({\tau}{|}{2i\lambda})}\over
{{\theta_1^\prime}({0}{|}{2i\lambda})}}\Biggr\rbrack^{-2}\,
{e^{-L^2\tau^2/2\pi\lambda}}{\cal C}
}
\noindent where
\eqn\cdef{\eqalign{
{{{\cal C}{\equiv}}}{{1}\over{n}}\oint_{z_1=0}&
{{dz_1}\over{2\pi}}\oint_{z_2=\tau}{{dz_2}\over{2\pi}}\cr
&\Biggl\lbrack{{{\theta_1}({\tau}-{z_1}|{2i\lambda)}{\theta_1}
({\tau}-{z_2}|{2i\lambda)}}
\over{{\theta_1}({z_1}|{2i\lambda)}{\theta_1}({z_2|2i\lambda)}}}\Biggr\rbrack
^n [{\ell n}{\theta_1}]^{\prime\prime} ({\tau}-{z_1}-{z_2}|{2i\lambda})
}}
\noindent and the extra factor of ${1/n}$ arises from normalizing the
DDF operators to unity.

Recall that we wish to investigate the expression as ${n},\,\,
{L}{\rightarrow}{\infty}$. One might have hoped this limit yields a
simple expression, as this is the apparently simple limit where one of
the branes is irrelevant and we simply have a plane wave propagating on
a semi-infinite string. Unfortunately, simplification does not obviously
occur. From the physical standpoint, other effects besides scattering
from the brane complicate matters. In particular, if one looks at the
annulus diagram in the closed string channel, it describes exchange of a
graviton between the brane and the now infinitely massive string.
Instead of analyzing the general expression, the next section will
consider it in the open-string factorization limit,
${\lambda}{\rightarrow} {0}$.

\newsec{Open String Factorization}

In order to investigate the expression \famp\ we will consider it in the
open string factorization limit, ${\lambda}{\rightarrow}{0}$.
In this limit it is convenient to use the modular transformation
properties of the theta functions to rewrite them as functions of 
\eqn\que{
{\tilde q} = {e^{-\pi/\lambda}}.
}
\noindent It is also convenient to define variables
\eqn\ses{
{s} = {e^{-\pi\tau/\lambda}}
}
\noindent and
\eqn\we{
{w_i} = {e^{\pi z_i/2\lambda}}\ .
}
The various contributions to \famp\ are readily expanded in terms of these
variables (as we'll see momentarily this leads to a direct physical
interpretation):
\eqn\currie{
\eqalign{
{{{\theta}({\tau}-{z_i}|{2i}{\lambda})}\over{{\theta}({z_i}|{2i}{\lambda})}}&=
{e^{-\pi\tau^2/2\lambda}}\,{w_i^{2\tau}}\,\,{\sqrt{{1}\over{s}}}\,\,
{{{w_i^{-1}}-{sw_i}}\over{{w_i}-{w_i^{-1}}}}\cr &{\prod_{n=1}^\infty}\,\,
{{({1}-{\tilde q}^n/{sw_i^2})({1}-{\tilde q}^n {sw_i^2)}}\over
{({1}-{\tilde q}^n/{w_i^2})({1}-{\tilde q}^n{w_i^2})}},\cr
[{\ell n\theta_1}]^{\prime\prime} ({\tau}-{z_1}-{z_2}|{2i\lambda})&=
{-}{{\pi}\over{\lambda}}-\biggl({{\pi}\over{\lambda}}\biggr)^2\,\,\Biggl[
{{s}\over
{({1/w_1}{w_2}-{sw_1}{w_2})^2}}\cr
&+\sum_{n=1}^\infty\,\,{{{\tilde q}^n/{sw_1w_2}}\over{({1}-{\tilde q}^n/
{sw_1w_2})^2}}\,\,+\,\,{{{\tilde q}^n{sw_1w_2}}\over{({1}{-}{\tilde
q}^n{sw_1w_2})^2}}\Biggr],\cr
}}
\noindent and
\eqn\gloria{
\biggl[{{{\theta_1}({\tau})}\over{{\theta_1^\prime}({0})}}\biggl]^{-2}\,\,{=}
{{\pi^2}\over{\lambda^2}}\,\,{e^{\pi\tau^2/\lambda}}\,\,
{{s}\over{({1}-{s})^2}}\,\,{\prod_{n=1}^\infty}\,\,
{{({1}-{\tilde q}^n)^4}
\over{({1}-{\tilde q}^n{s})^2\,({1}-{\tilde q}^n/{s})^2}}.
}
\noindent We also use
\eqn\vince{
2^{-12}{q^{-2}}{f^{-24}}({q^2}) = \lambda^{12}{\tilde q}^{-1}{f} 
({\tilde q})^{-24} = {\lambda}^{12}\,\,\sum_{\ell=0}^\infty
{d_\ell}{\tilde q}^{\ell-1},
}
\noindent where ${d_\ell}$ is the degeneracy of open string states at
level $\ell$.

\indent The preceding expressions imply the existence of an expansion of
${\cal C}$ in
${s}$ and ${\tilde q}$, of
the form
\eqn\leslie{\eqalign{
\biggl[{{{\theta_1}({\tau})}\over{{\theta_1^\prime}({0})}}\biggr]^{-2}\,\,
{\cal C} =&
{e^{-\pi\tau^2(n-1)/\lambda}}\cr 
&\sum_{m=0}^\infty\,\sum_{\ell=-m-n}^\infty\,\,
\biggl[{{2}\over{\pi\lambda}}\,\,{G_{\ell m}^1}({n}{\tau},{n}) + {{1}\over
{\lambda^2}}\,\,{G_{\ell m}^2} ({n}{\tau},{n})\biggr]\,\,{s^{\ell+1}}
{\tilde q}^m.}
}
\noindent This expression is readily interpreted after inserting into the
amplitude \famp,
\eqn\gertrude{
\eqalign{
{\cal A}&{=}
\sum_{r=0}^\infty\,{d_r}\,\sum_{m=0}^\infty\,\sum_{\ell=-m-n}^\infty\,
\int_0^\infty\,{d\lambda}\,\int_0^1\,{d\tau}\,{\lambda^{{p+1\over2}-3}}
\,\,\biggl[{{2\lambda}\over{\pi}}\,{G_{\ell m}^{1}}({n}{\tau},{n}) +
{G^2_{\ell m}} ({n}{\tau},{n})\biggr]\cr
&{\rm{exp}} \biggl\lbrace - {{L^2\tau^2}\over{2\pi\lambda}} -
{{\pi(n-1)\tau^2}\over{\lambda}} - {{\pi\tau}\over{\lambda}}\,({\ell+1})\,
- {{\pi}\over{\lambda}} ({m} + {r} - {1})\biggr\rbrace\ .
}} 
After defining ${T} = {{\pi}\over{2\lambda}}$, we find
\eqn\aamp{
\eqalign{
{\cal A}&{\propto}
\sum_{r=0}^\infty\,d_{r}\,\sum_{m=0}^\infty\,\sum_{\ell=-m-n}^\infty\,
\int_0^\infty\, {{dT}\over{{T^{(p+1)/2-1}}}}\int_0^{1}\,{d\tau}
\biggl[{1\over T} G_{\ell m}^1 ({n\tau,n}) + {G_{\ell m}^2}
({n\tau,n})\biggr]\cr
&{\rm{exp}} \left\{ - {T}{\tau^2} \biggl[{{L^2}\over{\pi^2}} +
{2} ({n} - {1})\biggr] - {2} ({\ell} + {1}) {T}{\tau} - {2} ({m} + {r} -
{1}) {T}\right\}\ .\cr
}} 
Ignoring the ${G_{\ell m}^i}$'s, this is the Schwinger proper
time
formula for the one loop amplitude of a particle with ${M^2} =
{{L^2}\over{\pi^2}} + {2}\,({n} - {1})$ splitting into two particles with
masses given by
\eqn\massf{
\eqalign{
{\mu_1^2}&= {2}\,({m} + {r} - {1})\cr
{\mu_2^2}&= {M^2} + {2}\,({m} + {r} + {\ell}).\cr
}}
\noindent The ${G_{\ell m}^i}$'s are effective coupling functions for
these decays.

The details of the kinematics are easily seen in the ${L}{\rightarrow}{\infty}$
limit. Recall from \sally, 
\eqn\trudi{
{M} = {{L}\over{\pi}} + {\omega}.
}
\noindent Similarly, \massf\ gives
\eqn\bobbi{
{\mu_2} = {M} + {{\pi}\over{L}} ({m} + {r} + {\ell}),
}
corresponding to a final state frequency for the long string  
\eqn\cathy{
{\omega} - {\Delta}{\omega}
}
\noindent with
\eqn\robbi{
- {\Delta}{\omega} = {{\pi}\over{L}} ({m} + {r} + {\ell}).
}
\noindent The threshold condition for cuts corresponding to decay is
${M} > {\mu_1} + {\mu_2}$, or 
\eqn\threshc{
{\Delta}{\omega} {\geq} {\sqrt{2\,(m+r-1)}} \ .
}
\noindent The interpretation of such cuts is clear: the infinite string
loses energy ${\Delta}{\omega}$ by emitting an open string at mass level ${m}
+ {r} $, with both of its ends attached to the brane.

There are three classes of such states. The first, the tachyonic state,
is an unphysical artifact and will be ignored. The massless states
correspond to elastic motion of the brane--either center of mass recoil
or tension-driven vibrations of the brane, along with world volume gauge field
excitations. For ${\omega} < {\sqrt{2}}$, \threshc\ shows that these are
the only modes excited. For ${\omega} > {\sqrt{2}}$, one also excites massive
string modes. These ``string halo" excitations 
may or may not be properly thought of as internal structural
degrees of freedom of the brane.

\newsec{High energy structure}

Although such ``halo'' modes have a threshold corresponding to the
string scale, one would like to know at what energy scale the 
scattering amplitudes become large. The relevance of this was alluded
to in the introduction: at ${\cal O} ({g})$ in the coupling constant,
the scattering has a constant phase shift, and thus appears sensitive to
the point-like structure of the D-brane. Including scattering effects at
${\cal O} ({g})$ and higher, we wish to determine the scale at which the
scattering becomes non-trivial. As suggested in the introduction,
power law behavior
\eqn\agrowth{
{{\rm Im}}\,{\cal A}\,{\propto}\,{g^2} {\omega^{2p}}
}
\noindent in the imaginary part of the one-loop amplitude would
suggest that the relevant scale is
${E} {\sim} {g^{-1/p}}$, exhibiting structure on sub-string distances.

A definitive answer to this question apparently requires determination
of the effective couplings ${G_{\ell m}^i}$. So far this has not been
possible. However, there is one clue to the large-${\omega}$ behavior of the
amplitude. At large ${\omega}$, the number of possible states for the
radiated string grows as ${e^{\sqrt{8} \pi\omega}}$, according
to standard arguments
regarding the asymptotic level density \refs{\GSW}.
This is contained in the explicit factor of ${d_{r}}$ in \aamp.
Therefore, unless the couplings ${G_{\ell m}^i}$  both decrease
exponentially rapidly as ${\omega}{\rightarrow}{\infty}$, the amplitude will
exhibit exponential behavior.

Of course, such a degeneracy factor occurs in other one loop string
amplitudes, and doesn't produce exponential enhancement. One example is
in high energy string scattering\refs{\GrMe,\Vene} where we see Regge behavior.
Another is in the case of decay of high-level leading Regge-trajectory
states \refs{\Bakl} where kinematical angular momentum constraints
forbid emission of all but certain restricted states. However, these
constraints don't apply to the present case of an initial state on a
highly subleading trajectory. Furthermore, the type of process being
considered is very different from high-energy string scattering.  In
high-energy string scattering, the softness of the string interaction
presumably interferes with efficient excitation of the exponential
degeneracy of available states -- the energy is not easily transferred to
world-sheet oscillations.  However, here one is 
starting out with a state that is high-frequency on
the worldsheet. This, and the effective hardness of the
D-brane interaction, may allow one to access the exponentially growing
number of states in this process.

Indeed, via a
rather lengthy analysis it is possible to find an approximation to
${G_{\ell 0}^i}$ that appears not to exhibit such exponential
suppression, although this analysis is not conclusive as it is not
inconceivable that cancellations with the
other ${G_{\ell m}^i}$'s could produce such suppression.

Therefore, it seems at least plausible that instead of \agrowth\ we have
\eqn\aguess{
{{\rm Im}}\,{\cal A}\,{\propto}\,{g^2}\,{e^{c\omega}}.
}
\noindent This would mean that the scale at which the scattering becomes
appreciable is the string scale, or at most logarithmically down from
the string scale. If this is the case, scattering ripples from branes
will not necessarily reveal the desired sub-stringy structure.

Confirmation of this (or the more interesting converse) requires the
application of more clever techniques to the expression \famp. Another
approach in the literature \refs{\Mitu} involves getting decay rates
from asymptotic analysis at {{\it large}} ${\lambda}$, but so far
\famp\ has not proven amenable to such treatment. Perhaps related or
other techniques will ultimately allow a check of the conjecture
\aguess.
\vskip .50truein
\noindent {\bf ACKNOWLEDGMENTS:}

I would like to thank J. Polchinski for interesting me in
this problem and for suggestions, and S. Chaudhuri for discussions.
Parts of this work was completed at the Aspen Center for Physics. This
work was supported in part by NSF PYI grant NSF-PHY9157463 and by grant 
DOE91ER40618.

\appendix{A}{}

This appendix contains a brief review/summary of some useful one-loop
technology.

\subsec{Green functions on annuli}

One needs explicit expressions for Green's functions on an annulus of
circumference 1 and height ${\lambda}$. First consider the case of the
torus. Since the function 
\eqn\mike{
{\theta_1} ({z}|{\tau}) = {i} \sum_{n=-\infty}^\infty ({-}{1})^n 
{q^{(n-1/2)^2}}\,{e^{i\pi(2n-1)z}}}
\noindent vanishes at ${z}={0}$, a natural guess for the field of a
point charge at ${w}$ is 
\eqn\jo{
{\ell}{n}\,{\theta_1} ({z} - {w}|{\tau}).}
\noindent However, this is not single valued under ${z}{\rightarrow}{z}
+ {1}$ and ${z}{\rightarrow}{z} + {\tau}$, under which

\eqn\matilda{
\eqalign{
{\theta_1} ({z} + {1}|{\tau})&= {-} {\theta_1} ({z}|{\tau}),\cr
{\theta_1} ({z} + {\tau}|{\tau})&= {-} {e^{-i\pi\tau-2i\pi
z}}\,\,{\theta_1} ({z}|{\tau}).\cr
}}
\noindent This is related to the fact that it is not possible to place a
single charge on a closed surface and obey Gauss's law. The solution is
to introduce a background charge so that the net charge is zero. There
are many ways to do this, but one is simply to put the background charge at
a point, 
${z} = {w'}$. One still has to add a homogeneous solution of
Laplace's equation to satisfy all periodicity conditions, and the resulting
(chiral) Green function is 
\eqn\Tgf{
{G_T} ({z};{w},{w^\prime}|{\tau}) = {{1}\over{2}}\,{\ell}{n}\,\,
\biggl[{{{\theta_1} ({z} - {w}|{\tau})}\over{{\theta_1} ({z} -
{w^\prime}|{\tau})}}\biggr]\,\,{-} {{{\pi}{i} ({w} -
{w^\prime}){\rm Im}{z}}\over{\rm Im}{\tau}}.
}
Combining this with a similar expression involving $\zbar, \wbar$, \etc,
gives the full scalar Green function with the normalization convention $\nabla^2
G = 2\pi \delta$.  An important consistency check on amplitudes (momentum
conservation) is that dependence on the arbitrary point
${w^\prime}$ drops out.

The Green functions for the annulus are easily constructed using the
fact that the double of the annulus, with modulus ${\lambda}$, along its
boundaries is the torus with modulus ${\tau} = {2i\lambda}$. Thus, we
simply place image charges to satisfy the appropriate boundary
conditions. For Neumann boundary conditions, $({\partial_z} -
{\partial}_{\bar z}) {G_N} ({z},{w},{w^\prime}|{\lambda}) =
{0}$ at ${\rm Im}{z} = {0}, {\lambda}$ is satisfied by
\eqn\melonie{
{G_N} ({z};{w},{w^\prime}|{\lambda}) = {{\ell}{n}}\,\Big|
{{{\theta_1} ({z} - {w}|{2i\lambda}) {\theta_1} ({z} -
{\bar w}|{2i\lambda})}\over{{\theta_1} ({z} - {w^\prime}|{2i\lambda})
{\theta_1} ({z} - {{\bar w}^\prime}|{2i\lambda})}}\Big|,
}
\noindent which is easily seen to have the required periodicity under
${z}{\rightarrow}{z} + {1}$. For Dirichlet boundary conditions ${G_D}
({z},{\omega}|{\lambda}) = {0}$ at ${\rm Im}{z} = {0}, {\lambda}$, here
no background charge is needed, and we find
\eqn\stacy{
{G_D} ({z},{w}|{\lambda}) = {\ell}{n}\,\Big|
{{{\theta_1} ({z} - {w}|{2i\lambda})}\over{{\theta_1} ({z} -
{\wbar}|{2i\lambda})}}\Big| {-} {{2\pi i (w - \wbar)\rm{Im}{z}}\over
{{\rm Im}{\tau}}}\ .
}
\subsec{Functional determinants on annuli}

The determinant of ${\Delta} = {-} {\nabla^2}$ on an annulus with
either Neumann or Dirichlet boundary conditions is easily written in
terms of evaluable infinite products. With Neumann boundary conditions,
eigenfunctions are 
\eqn\rachel{
{\chi}_{m,n} = {e^{2\pi i n \sigma^1}}\,\,{\cos}
\biggl({{m\pi}\over{\lambda}} {\sigma^2}\biggr)\ ,
}
\noindent and with Dirichlet boundary conditions, 
\eqn\lisa{
{\psi}_{m,n} = {e^{2\pi i n\sigma^1}}\,\,{\sin}
\biggl({{m\pi}\over{\lambda}} {\sigma^2}\biggr).
}
These yield
\eqn\julie{
{\rm det}_N^\prime{\Delta} = \biggl ({\prod_{n=1}^\infty}\,{4}{\pi^2}{n^2}
\biggr)^2 {\prod_{m=1}^\infty}\,\,{{m^2\pi^2}\over{\lambda^2}}\,\,
{\prod_{n=1}^\infty} \biggl( {{m^2\pi^2}\over{\lambda^2}} +
{4}{\pi^2}{n^2}\biggr)^2
}
and
\eqn\flower{
{\rm det}_D{\Delta} = {\prod_{m=1}^\infty}
{{m^2\pi^2}\over{\lambda^2}} {\prod_{n=1}^\infty} \biggl(
{{m^2\pi^2}\over{\lambda^2}} + {4}{\pi^2}{n^2}\biggr)^2,
}
where prime indicates omission of the zero mode. These
infinite products are readily evaluated using identities given in, for
example, \refs{\SGTasi}.

\noindent In particular,
\eqn\helen{
{\prod_{n=1}^\infty}\,{a} = {{1}\over{\sqrt{a}}}\ ,
}
for arbitrary constant $a$, and 
\eqn\bobbie{
{\prod_{n=1}^\infty}\,{n^2} = {2\pi},
}
\noindent so the two determinants are equal. Further application of such
identities (see \SGTasi) gives
\eqn\eve{
{\rm det}_N^\prime {\Delta} = {\rm det}_D{\Delta} = 2{\lambda}\Biggl[
{e^{-{{\pi\lambda}\over{6}}}}\,\,{\prod_{n=1}^\infty} \biggl({1} -
{e^{-4\pi\lambda n}}\biggr)\Biggr]^2.
}

\listrefs


\end